# MASSES AND IMPLICATIONS FOR AGES OF LOW-MASS PRE-MAIN SEQUENCE STARS IN TAURUS AND OPHIUCHUS

M. Simon[1,2], S. Guilloteau[3], Tracy L. Beck[4], E. Chapillon[3,7], E. Di Folco[3], A. Dutrey[3], Gregory A. Feiden[5] N. Grosso[6],V. Piétu[7]. L. Prato[8], Gail H. Schaefer[9]

[1]Dept. of Physics and Astronomy, Stony Brook University, Stony Brook, NY 11794-3800, USA; michal.simon@stonybrook.edu

[2]Dept. of Astrophysics, American Museum of Natural History, New York, NY 10024, USA

[3]Laboratoire d'Astrophysique de Bordeaux, Univ. Bordeaux, CNRS, B18N, allée Geoffroy Saint-Hilaire, 33615 Pessac, France

[4]Space Telescope Science Institute, 3700 San Mateo Drive, Baltimore, MD 20218, USA

[5]Dept. of Physics and Astronomy, University of North Georgia, Dahlonega, GA, 30597, USA

[6]Aix Marseille University, CNRS, CNES, LAM, Marseille, France

[7]IRAM, 300 rue de la piscine, F-38406 Saint Martin d'Hères, France

[8]Lowell Observatory, 1400 West Mars Hill Road, Flagstaff, AZ 86001, USA

[9]The CHARA Array of Georgia State University, Mount Wilson Observatory, Mount Wilson, CA 91023, USA

## ABSTRACT

The accuracy of masses of pre-main sequence (PMS) stars derived from their locations on the Hertzsprung-Russell Diagram (HRD) can be tested by comparison with accurate and precise masses determined independently. We present 29 single stars in the Taurus star-forming region (SFR) and 3 in the Ophiuchus SFR with masses measured dynamically to a precision of at least 10%. Our results include 9 updated mass determinations and 3 that have not had their dynamical masses published before. This list of stars with fundamental, dynamical masses, $M_{dyn}$, is drawn from a larger list of 39 targets in the Taurus SFR and 6 in the Ophiuchus SFR. Placing the stars with accurate and precise dynamical masses on HRDs that do not include internal magnetic fields underestimates the mass compared to $M_{dyn}$ by about 30%. Placing them on an HRD that does include magnetic fields yields mass estimates in much better agreement with $M_{dyn}$, with an average difference between $M_{dyn}$ and the estimated track mass of $0.01 \pm 0.02$ $M_\odot$. The ages of the stars, 3–10 MY on tracks that include magnetic fields, is older than the 1–3 MY indicated by the non-magnetic models. The older ages of T Tauri stars predicted by the magnetic models increase the time available for evolution of their disks and formation of the giant gas exoplanets. The agreement between our $M_{dyn}$ values and the masses on the magnetic field tracks provides indirect support for these older ages.

## 1. INTRODUCTION

The mass of a star is its most fundamental parameter and governs its life from formation to end. In broad outline, astronomers traditionally determine the masses of stars in two ways: by location on the HRD with reference to theoretical models of stellar evolution, and gravitationally by analyzing the orbits of binary stars. HRD-fitting is reliable for stars on the main sequence where the accuracy of the theoretical models has been well established. In fact, the HRD-fitting approach on the main sequence owes its success to the many masses of binary stars determined from their orbital motion.

The HRD is the fundamental tool for the analysis of main sequence (MS) stars and of their evolution away from the MS because it provides a template on which large numbers of stars can be studied together. Astronomers therefore also want to use the HRD to determine the masses and ages of very young stars in clusters and associations where star and planet formation are occurring. However, divergence among the theoretical models has been an obstacle to performing this reliably. Frustratingly, differences among the models have been the most pronounced for stars of greatest interest in star forming regions, masses < 1.5 $M_\odot$ and age < 10 MY (Simon et al. 2013). As a result, astronomers' understanding of the earliest stages of star and planet formation is incomplete. This is a serious limitation because at least one planet associated with a PMS star, CI Tau b, is already known (Johns-Krull et al. 2016; Flagg et al. 2019). However, there is now much greater agreement among recent models of PMS evolution. For stellar masses $> \sim 0.5$ $M_\odot$ and age $> 1 Myr$ the models of Baraffe et al. (2015), *Modules for Experiments in Astrophysics, MESA* (Paxton et al. (2015), Choi et al. (2016)) and the non-magnetic models of Feiden (2016), are in excellent agreement (Simon & Toraskar 2017).

To provide reliable masses for the theoretical models we have been engaged in an active effort to measure masses of PMS stars in visual binaries (VBs), spectroscopic binaries (SBs), and single stars. Very young single stars offer a unique dynamical approach to mass measurement because many stars younger than $\sim 3$ MY are surrounded by rotating gaseous and dusty circumstellar disks. These stars are the Class II T Tauri stars. The techniques are complementary in the sense that mass measurements by the circumstellar disk technique are necessarily limited to the Class II T Tauris while the VB and SB techniques may be applied to Class II and Class III T Tauris.

The mass of a star surrounded by circumstellar disk can be measured in a single interferometric observation while measurement of the orbits of VBs and SBs in the nearest SFRs can require many allocations of telescope time and take years. The number of PMS stellar masses determined by the circumstellar disk technique is therefore larger than by measurements of VBs and SBs. To set the physical scale of the disks measured interferometrically, their distances are needed. Precise distances are now available to most PMS stars in the Taurus SFR using the second release (DR2) of the *GAIA* mission[1]. In this paper we apply the distances in DR2 to the PMS stars with masses measured by the circumstellar disk technique. Our sample consists of stars with previously measured dynamical masses and also new determinations derived using archival ALMA data. Details are given in §2. Our results provide dynamically determined fundamental values of the mass, $M_{dyn}$. These masses are fundamental because they are independent of any other stellar property. We define a subset of the sample that has $M_{dyn}$ known to at least 10%. In §3 we plot this subset on HRDs that show PMS evolutionary tracks and isochrones calculated with and without internal magnetic fields in order to compare dynamical masses with those derived with respect to the tracks, $M_{dyn}$ and $M_{hrd}$s, respectively. For clarity and simplicity we will refer to the evolutionary models without magnetic fields as standard models and those that do include them as magnetic models.

## 2. SAMPLE AND DATA

### 2.1. *Overview: The Samples in the Taurus and Ophiuchus SFRs*

Our sample concentrates on objects in the Taurus SFR because the most data are available in this region. For completeness, however, we include in our analysis the stars in the Ophiuchus SFR that we reported on previously (Simon et al. 2017). The map of the Taurus SFR (Fig. 1) shows all the PMS stars identified by Luhman (2018) that have DR2 distances provided by Bailer-Jones et al. (2018). The map contains 310 stars but not all can be distinguished given the size of the plotting symbols. Distances are indicated by the color-coding. Most of the stars lie at a distance near 140 pc as expected (Kenyon et al. 1994) but are distributed over distances 100 to 200 pc. Fig. 1 also shows that stars tend to lie at distances greater than 140 pc toward less negative galactic latitudes. Interestingly, several regions contain at least one star that lies at a statistically significantly greater distance than its apparent neighbors on a 2-D representation of the region.

Table 1 lists the 45 stars in Taurus in our sample. The stars are singles and components of binaries that are sufficiently wide that the disk rotation of one component can be measured without confusion from the other. The sample includes 33 stars with previously published dynamical masses and 12 stars for which we obtained new dynamical masses using archival ALMA data (§2.2). Of the 12 stars, 5 have not had dynamical masses published previously. Col. 1 in Table 1 lists the stars by name in order of right ascension. Cols. 2 and 3 give their galactic longitude and latitude; we provide the galactic coordinates to facilitate locating the stars in Fig. 1. They are not intended for purposes of observing the stars. The previously published masses, $M_{pub}$, and distances at which they were evaluated, $D_{pub}$, are given in Cols. 4 and 5; Col. 8 provides the references for these values. Col. 6 lists Bailer-Jones et al. (2018)'s Gaia DR2 distance estimates of these stars. Col. 7 lists the stellar masses at the DR2 distances. The uncertainties in $M_{dyn}$ include those of the DR2 distance estimate. Stars for which a DR2 distance estimate is not available have their measured mass listed in Col. 4 at the 140 pc average distance to the Taurus SFR. Gaia DR2 distances are available for 35 stars in Table 1; they are indicated in Fig. 1 by overplotted squares. Of 35 fundamental masses, 3 are determinations for stars not previously published; these are derived from archival ALMA data. Another 9, also from archival data, are new determinations for stars previously published. The remaining 26 fundamental masses are for stars previously published and here placed at their Gaia DR2 distances.

---

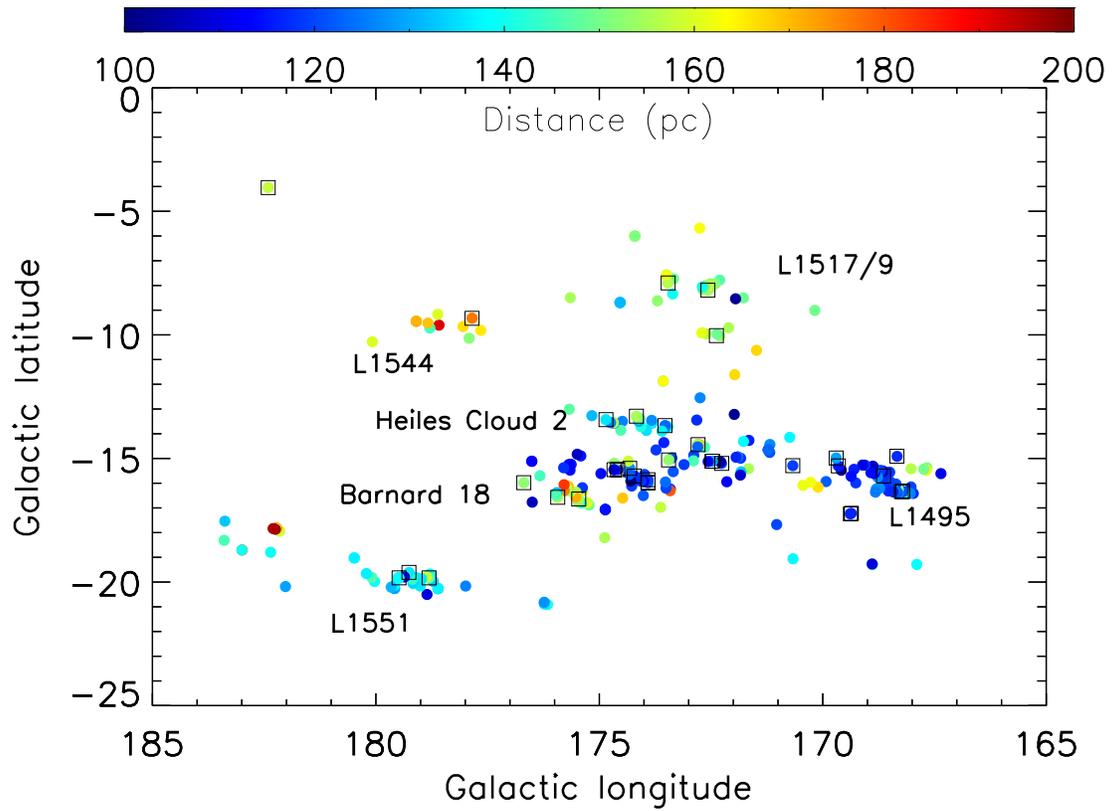

**Figure 1.** The colored circles represent the PMS stars in the Taurus SFR identified by Luhman (2018) with Gaia DR2 distances color-coded (in pc) as given by the colorbar. The 35 stars in Table 1 that have measured dynamical masses and Gaia DR2 distances are represented by superposed boxes (see §2.1).

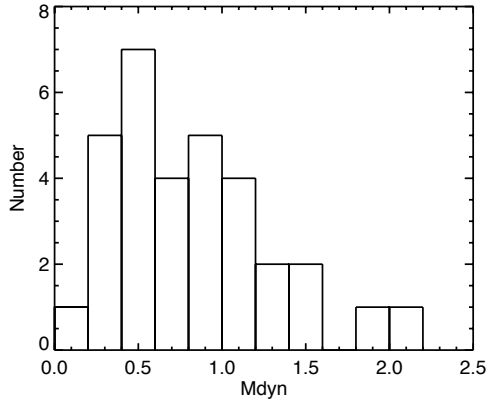

**Figure 2**. Distribution by number of $M_{dyn}$ in units of $M_\odot$, with uncertainties $< 10\%$ as listed in Tables 1 and 2.

Table 2 pertains to previously published masses (Simon et al. 2017) measured in the heavily obscured L1688 region of the Ophiuchus SFR. Its format is identical to Table 1 for Taurus. The 6 stars in the Ophiuchus region are fainter than the stars in Taurus owing to their obscuration and DR2 distances are available for only 3 of the stars in our sample.

### 2.2. Analysis of ALMA Archival Data

The archival data were calibrated using the appropriate CASA software for each observation, and then exported through UVFITS format to the GILDAS package for imaging and data analysis. For each disk observed in a CO line, we used our DiskFit package (Piétu et al. 2007) as in Simon et al. (2017) to fit a standard, vertically isothermal, power-law disk model to the data, with the addition of a CO depletion zone near the mid-plane. No CO was allowed to exist at points whose (r,z) cylindrical coordinates are such that $z/r < a_{dep}$, mimicking CO depletion on grains near the disk plane. The adjustable parameter $a_{dep}$ was typically found to be small ($< 0.1$) or negligible in all sources. The apparent scale height of the CO distribution was a free parameter in our model, since CO can be optically thick and collisionally excited up to 3-4 hydrostatic scale heights. The fit is performed by minimizing the $\chi^2$ between the modeled and the observed visibilities. The use of power laws limits our ability to accurately fit the brightness distribution as a function of radius. However, this has a negligible impact on the derived kinematics: for example, changing the temperature by 20 % or the CO column density by a factor 2 would not affect the derived velocities, and hence the derived stellar mass.

Significant contamination by the surrounding molecular cloud is present in most sources, so a range of channels around the cloud systemic velocity was ignored in the fitting process. We verified that our derived masses were not sensitive to the precise range of the avoided channels. Furthermore, we found that contamination by the molecular cloud in the CO J=3-2 line is much less significant than in the J=2-1 transition used in other studies. This is probably a result of insufficient excitation of the J=3-2 transition in the molecular clouds because of their low densities and temperatures. In a few cases, we relied on the $^{13}$CO isotopologue because of source complexity (CQ Tau, CW Tau) or lack of other tracer as in AA Tau (Appendix Fig. 1). We also analyzed the continuum emission in order to verify the consistency of the derived positions, orientation and inclinations. The Appendix provides details pertaining to each newly analyzed star and Table 1 in the Appendix lists the derived disk parameters and dynamic masses.

### 2.3. Fundamental Masses of Stars with Precisions $\leq 10\%$ and their Stellar Parameters

We selected for further analysis only the 32 stars with fundamental dynamical masses determined to relative uncertainty less than 10%; of these, 29 are in Taurus and 3 in Ophiuchus. The stars and their values of $M_{dyn}$ are listed in Columns 1 and 2 of Tables 3 and 4. Fig. 2 shows the mass distribution of the 32 stars. Our sample contains few stars with $M_{dyn} > 1M_\odot$ because early in our program we chose to concentrate on PMS stars with expected values of $M_{dyn}$ less than $\sim 1M_\odot$ where the discrepancies among the theoretical models were the greatest. Only 6 stars in our sample have $M_{dyn} \leq 0.4\ M_\odot$. This situation arises because we found empirically that observable angularly resolved disks around the lowest mass PMS stars are rare. Clearly the sample needs to be increased in this mass range in order to provide reliable mass references for models of PMS evolution.

Table 3 gives the photospheric parameters for the stars in the Taurus SFR. The spectral type in Col. 3 is quoted from Luhman (2018). The table presents two sets of parameters with the published luminosities scaled to DR2 distances.

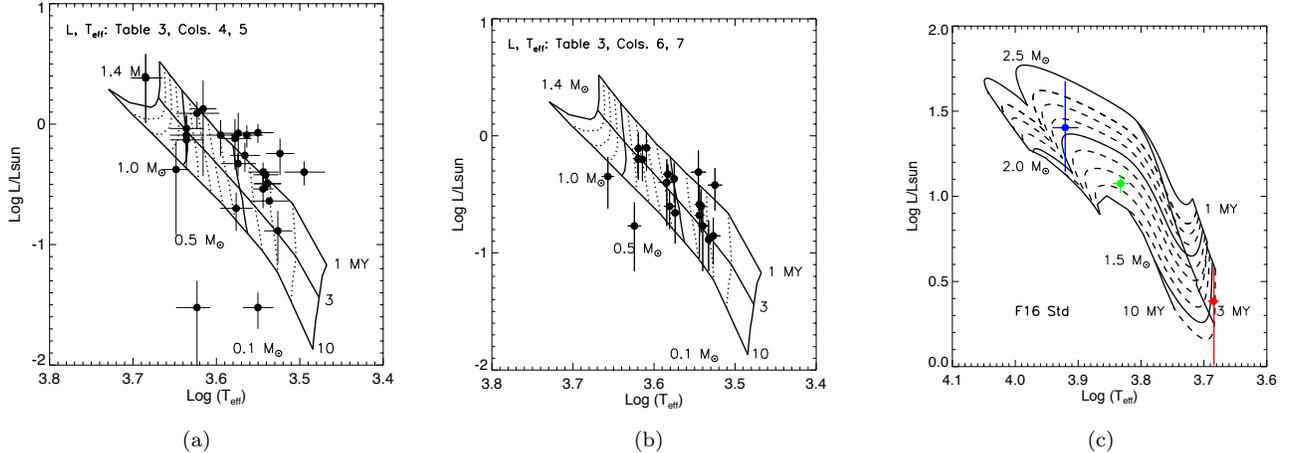

**Figure 3.** *a and b:* PMS stars in Taurus (Table 3) on Hertzsprung-Russell Diagrams using the evolutionary models of Baraffe et al. (2015). *c:* The HAe/Be stars MWC 480 (red) and CQ Tau (green) and the T Tauri CW Tau (blue) on an HRD using the standard models of Feiden (2016).

The luminosities in Col. 4 are from Andrews et al. (2013), except that of CQ Tau which is from Villebrun et al. (2019). Col. 5 lists the effective temperatures, $T_{eff}$s, corresponding to the spectral types in Col. 3 using Pecaut & Mamajek (2013)'s look-up table. The uncertainties correspond to one spectral type subclass. Andrews et al. (2013) derived the luminosities from a best fit to the stellar $T_{eff}$, extinction, and spectral energy distribution extending to millimeter wavelengths. The second set of parameters uses the luminosities (col. 6) of Herczeg & Hillenbrand (2014) with 0.2 dex uncertainties (G. Herczeg, priv. comm.). The $T_{eff}$'s use spectral types given by Herczeg & Hillenbrand (2014) with the conversion given in their Table 5. Herczeg & Hillenbrand (2014) calculated luminosities by applying the observed visible light spectrum to a model atmosphere at the appropriate $T_{eff}$ and extinction to derive a photospheric luminosity and scaling this to the stellar luminosity with a bolometric correction derived from the spectrum.

Table 4 pertains to the stars in the Ophiuchus SFR. The luminosities and spectral types in Columns 3 and 4 for GSS 39 and YLW 58 are from Ricci et al. (2010) and those for ROX 25 (GY 292, 2MASS J16273311-2441152) are from Erickson et al. (2011). The parameters in Columns 6 and 7 are from Najita et al. (2015). In all cases the effective temperature conversion uses the look-up table of Pecaut & Mamajek (2013) as above.

## 3. RESULTS AND DISCUSSION

### 3.1. *Track Masses from Hertzsprung-Russell Diagrams*

To compare the mass of a star that would be derived from its $(T_{eff}, L)$ position on the HRD with it actual mass $M_{dyn}$ we place the stars in Tables 3 and 4 on HRDs that use both standard models (Figs. 3 and 4) and magnetic models (Fig. 5) of PMS evolution. Figs 3a and b use the standard models of Baraffe et al. (2015). Fig. 3a plots the stellar parameters in Cols. 4 and 5 of Table 3 and Fig. 3b plots the parameters in Cols. 6 and 7. Fig. 3a shows that CW Tau, despite its value of $M_{dyn} = 0.64 \pm 0.01$, lies above the highest mass track, 1.4 $M_\odot$, available in Baraffe et al. (2015)'s models. Therefore we plot CW Tau and the more massive HAe/Be stars MWC 480 and CQ Tau on an HRD (Fig. 3c) which uses the standard models of Feiden (2016). The HRD locations of most of the stars in Figs. 3 indicate ages between 1 and 10 MY, with the majority between 1 and 3 MY. Uncertainties in the age are dominated by the uncertainties in the luminosities. The relatively low luminosities of HK Tau B and HV Tau C indicate implausible ages > 10 MY. Their luminosities are probably underestimated because the stars are observed through disks that are nearly edge-on (Appendix, Table 1). The distributions of L and $T_{eff}$ in Figs. 3a and b are similar, but as a group the luminosities derived by Herczeg & Hillenbrand (2014) are smaller than those derived by Andrews et al. (2013) by about 0.3 dex.

Fig. 4 shows the HRD for the 3 stars in Ophiuchus for which we could evaluate their $M_{dyn}$. Their $(T_{eff}, L)$ values suggest ages < 1 MY except for ROX 25. Its high luminosity given by Najita et al. (2015) suggests an age $\ll$ 1 MY but Erickson et al. (2011)'s luminosity estimate corresponds to age $\sim$ 5 MY. The stellar luminosities and hence ages in L1688 are very uncertain because of the high extinction to the stars.

We evaluated a star's track mass $M_{hrd}$ in the traditional way by its position relative to the evolutionary tracks. We list these values in Table 5, $MA_{hrd}$ for the stars in Taurus in Fig. 3a and $MB_{hrd}$ using Fig. 3b. The uncertainty is derived by drawing an ellipse centered at $(logT_{eff}, logL)$ with shape given by uncertainties in luminosities scaled to DR2 distances. We could not do this reliably for the 3 stars in Ophiuchus plotted in Fig. 4 because of their uncertain luminosities.

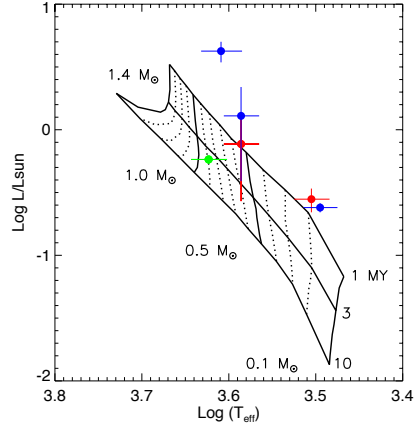

**Figure 4.** PMS stars in Ophiuchus on HRDs using Baraffe et al. (2015)'s models. The red dots plot GSS 39 and YLW 58 using Ricci et al. (2010)'s parameters and the green dot represents the parameters of ROX 25 (Erickson et al. 2011) (Table 4, Cols 4 and 5). The blue dots show the parameters of Najita et al. (2015) (Table 4, Cols. 7 and 8).

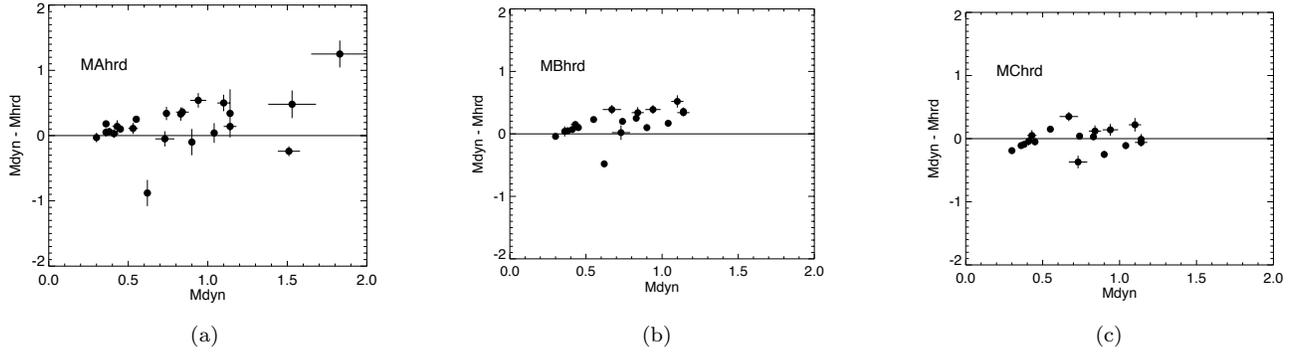

(a)                           (b)                           (c)

**Figure 5.** Values of $MA_{hrd}$ and $MB_{hrd}$ (Table 5) were derived using the two sets of $(T_{eff}, L/L_{\odot})$ estimates listed in Table 3. The $MC_{hrd}$ values (Table 5) were derived using the same stellar parameters as the $MB_{hrd}$ but placed on the HRD that uses magnetic evolutionary tracks. (a:)($M_{dyn} - MA_{hrd}$) vs $M_{dyn}$ and (b:) ($M_{dyn} - MB_{hrd}$ vs $M_{dyn}$. (c:) ($M_{dyn} - MC_{hrd}$) vs $M_{dyn}$. All masses are in units of $M_{\odot}$.

Fig. 5a shows that of the 25 stars with $MA_{hrd}$ values the track masses are smaller than the dynamical masses for 20. Only 2 stars have $M_{hrd} > M_{dyn}$ at more than the $3\sigma$ level, the classic T Tauri CW Tau and the HAeBe star CQ Tau. The $MB_{hrd}$ values are qualitatively similar as shown in Fig. 5b. The average difference ($M_{dyn}$ - $MA_{hrd}$) is $0.11 \pm 0.04$ $M_{\odot}$ and for $MB_{hrd}$ the average difference is $0.19 \pm 0.02$ $M_{\odot}$. In the $M_{dyn}$ mass range 0.4 to 1.0 $M_{\odot}$, the normalized differences ($M_{dyn}$ - $MA_{hrd}$)/$M_{dyn}$ and ($M_{dyn}$ - $MB_{hrd}$)/$M_{dyn}$ are about 0.3 if CW Tau and CQ Tau are ignored. These results followed from comparing the $(T_{eff}, L)$ positions of stars with respect to the models of Baraffe et al. (2015) but they would have been the same if we had used the standard models of Feiden (2016) or those calculated using MESA (Paxton et al. 2015). Evidently masses of PMS stars derived from the HRD that uses standard evolutionary models tend to underestimate the dynamically measured values. Hillenbrand & White (2004) obtained the same result using models of PMS evolution, all non-magnetic, available at the time.

The track mass depends mostly on the star's $T_{eff}$ because, at these ages, the evolutionary tracks are nearly vertical in the HRD. Apparently the $(T_{eff}, L)$ location of a star relative to a standard model of PMS evolution places it on a track of lower mass than its actual mass $M_{dyn}$. Such a track has a lower $T_{eff}$ suggesting that real stars are cooler than the standard model predictions. Fig. 6 (left) illustrates this. The solid lines indicate the theoretical values of $T_{eff}$ on the standard model *vs* stellar mass (Baraffe et al. 2015) at ages 1, 3, and 5 MY. The dots designate the $T_{eff}$ indicated by spectral types of the stars plotted in Fig. 3b. Regardless of a star's actual age, the stars' $T_{eff}$ values lie about 200-300 K below the theoretical values indicated by the non-magnetic model. This could be the result of an increased photospheric radius attributable to magnetic pressure in the interior or to significant starspot coverage of their photospheres, and in actuality, probably both (Somers & Pinsonneault 2015; Feiden 2016; Gully-Santiago et al. 2017; MacDonald & Mullan 2017).

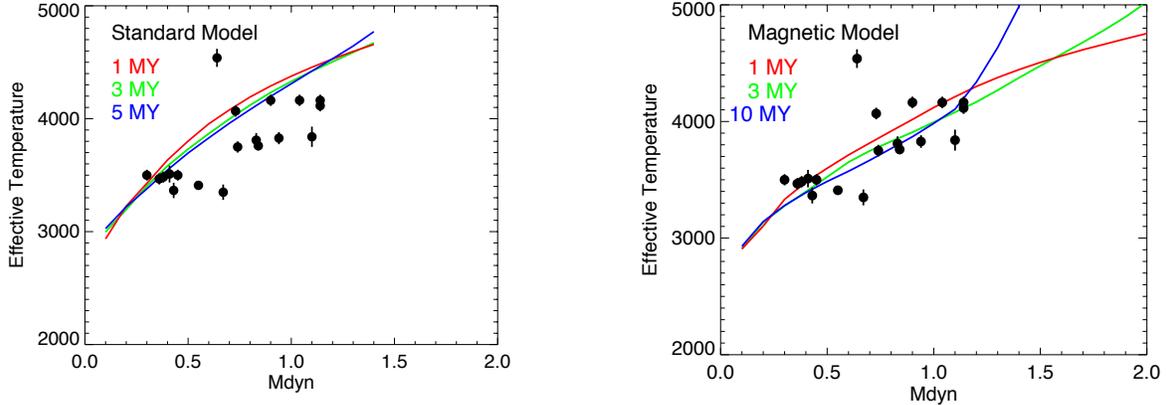

**Figure 6.** The solid lines are $T_{eff}(theory)$ vs stellar mass as given in the Baraffe et al. (2015) models (*left*) and Feiden's magnetic models (*right*). The data points are values of $T_{eff}$ derived using Herczeg & Hillenbrand (2014)'s stellar parameters for stars with the indicated $M_{dyn}$ ($M_\odot$).

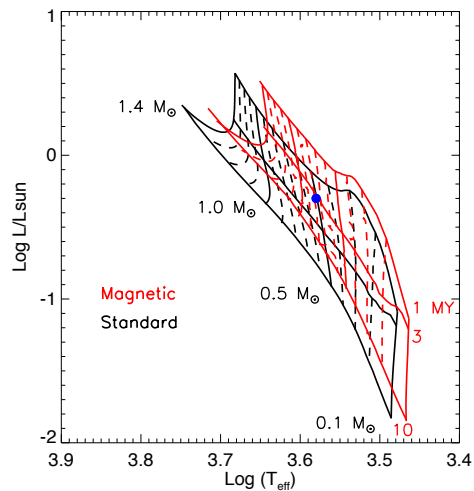

**Figure 7.** Comparison of Feiden's standard and magnetic models on the HRD.

To consider the effects of magnetic evolutionary models on the $M_{hrd}$ we used the models of Feiden (2016). In these models the magnetic field is in pressure equilibrium with the gas pressure at Rosseland mean optical depth = 1. The models are available from 0.1 MY to 10 GY; here we use them only in the age range 1 to 10 MY. Fig. 7a compares the magnetic and standard models. The blue dot represents a PMS star at $\log (T_{eff}) = 3.58$ and $\log (L/L_\odot) = -0.30$. Its mass and age measured with respect to the standard model would be $\sim 0.5$ $M_\odot$ and $\sim 2$ MY while relative to the magnetic model its mass and age would be greater $\sim 0.7$ $M_\odot$ and older $\sim 3$ MY. The older age is the result of the star's slower contraction owing to the additional magnetic pressure. It takes longer for the star to collapse to the photospheric radius determined by its luminosity and effective temperature. Further, in the presence of the added combined gas and magnetic pressures, a stronger gravitational field than in the standard model is necessary to support the star at the given photospheric radius. This requires a larger stellar mass.

Fig. 8 (left) shows the stars in Cols. 6 and 7 of Table 3 plotted on the HRD with the magnetic evolutionary tracks and isochrones. As expected, the stars appear older, roughly 3 to 10 MY, than in the standard model. We used this HRD to evaluate the track masses, $MC_{hrd}$, as before and list them in Table 6. The values of $MC_{hrd}$ exceed those derived using the standard models. Fig. 5c shows their distribution with $M_{dyn}$; the average difference ($M_{dyn}$ - $MC_{hrd}$) is $0.01 \pm 0.02$ $M_{\odot}$. Evidently the magnetic models provide mass estimates that are more consistent with the dynamical mass, $M_{dyn}$, than the standard models. Also, Fig. 6b shows that theoretical $T_{eff}$s of the magnetic models are in much better agreement with the $T_{eff}$s given by the spectral types than is the case in Fig. 6a for the standard model. However, the scatter of the observed $T_{eff}$s around the theoretical $T_{eff}$, in the plausible age range 1-10 MY, far exceeds their uncertainties. CW Tau at $M_{dyn} = 0.64$ $M_{\odot}$ and V710 Tau at $M_{dyn} = 0.69$ $M_{\odot}$ are extreme examples of stars that are hotter and cooler than the magnetic models predict. There are 3 possible reasons for this. Hot spots attributable to accretion will increase the apparent $T_{eff}$s while dark spots will lower it. Also the models may need to be adjusted for a star that has particularly strong or weak convection and hence correspondingly strong or weak internal magnetic field. Indeed, CW Tau has particularly strong mass accretion (Muzerolle et al. 1998). However, BP Tau which has a lower $T_{eff}$s than the models predict for its $M_{dyn}$ and would suggest spot coverage includes both accretion activity and dark spots (Donati et al. 2008). T Tauri stars often reveal both mass accretion through their accretion luminosity and star spots through periodic photometric variability. Separating these effects may have to proceed separately for each star.

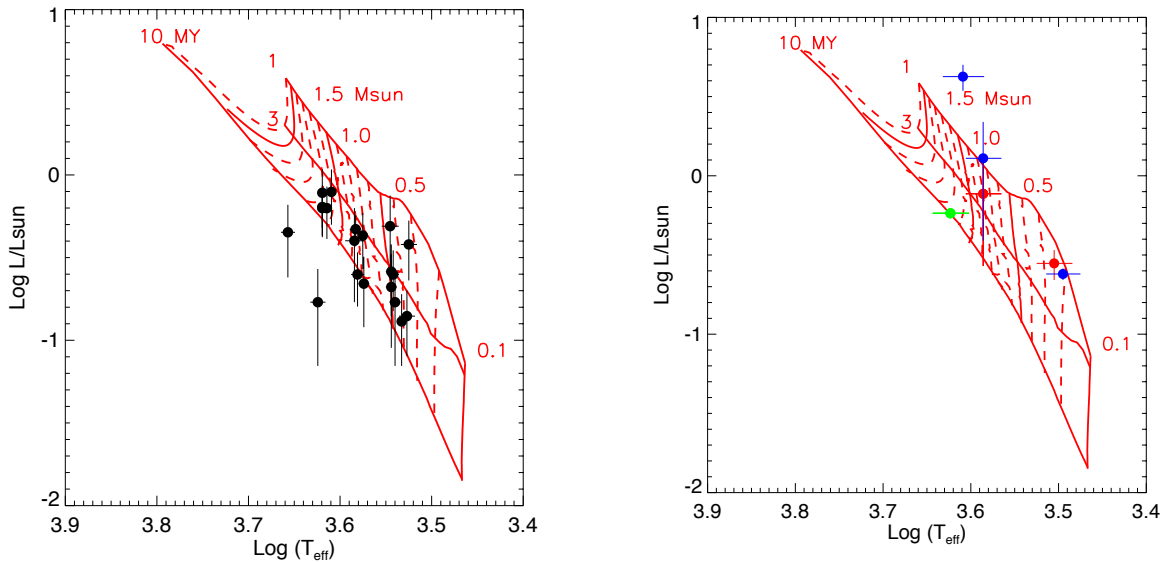

**Figure 8.** HRDs that use Feiden's magnetic models for Taurus *(left)* and Ophiuchus *(right)*. The data used are the same as in Figs 3b and 4. The color coding in the Ophiuchus figure is the same as in Fig. 4.

Fig. 8b shows the stars in Ophiuchus on the HRD using the magnetic models. The stars and their color-coding are the same as in Fig. 4. On this HRD ages of the stars are mostly between 1 and 3 MY. We evaluated their track masses $MD_{hrd}$ and $ME_{hrd}$ for the stellar parameters of Ricci et al. (2010), Erickson et al. (2011), and Najita et al. (2015) and list these in Table 6. We do not regard these track masses as reliable owing to the uncertainties of determining luminosities and spectral types of these very obscured stars.

### 3.2. *Implications of Older Ages for Exoplanet Formation and Disk Evolution*

Astronomers have drawn attention to the likelihood that PMS stars are several times older than implied by the standard evolutionary models by considering their Li-depletion ages (Jeffries (2017), Jeffries et al. (2017)), color-magnitude diagrams of young clusters (Mayne & Naylor (2008), Naylor (2009), Bell et al. (2013)), application of magnetic models to the components in eclipsing binaries with well-determined radii and masses (MacDonald & Mullan (2017), David et al. (2019)), and stars with parameters determined by photometry, spectroscopy and atmospheric models (Malo et al. 2014). Our finding that the dynamical masses of PMS stars in Taurus are consistent with ages 2–3 times older than the canonical 1–3 MY strengthens these results and supports the validity of applying the magnetic models to PMS stars.

The older ages of T Tauri stars predicted by the magnetic evolutionary tracks have significant implications for our understanding of planet formation and proto-planetary disk evolution. The older ages imply that gaseous circumstellar disks survive longer than currently expected. This leaves more time for chemical evolution, in particular on the dust grain surface, which could play a role in the apparent depletion of C and O (Favre et al. 2013) through, for example conversion of CO to $CO_2$ on dust grains ([Reboussin et al. 2015](#)) or long carbon chains. The older ages leave more time for giant planets to accrete their gaseous content. Current models have to rely on planet migration and a reduced dust opacity to be able to build giant gaseous planets within the 1-3 MY timescale previously accepted as the typical lifetime of disks (e.g., [Papaloizou & Nelson 2005](#); [Lissauer 2009](#)). Relaxing the time constraint by a factor 3 provides more flexibility in building the very diverse planetary systems that are observed.

## 4. SUMMARY

The main result of this work is the identification of a sample of stars with accurate and precise masses and comparison of these stars' dynamical masses with masses measured from models of PMS evolution to assess the accuracy of mass estimates from the HRD only. To this end, we culled our sample of 39 stars in the Taurus SFR and 6 in the Ophiuchus SFR with masses measured by rotation of their circumstellar disk to 29 stars in Taurus and 3 in Ophiuchus, with dynamical mass, $M_{dyn}$, good to at least 10%. We placed these on HRDs using models of PMS evolution with and without internal magnetic fields.

We found that:

1) In the mass range $\sim 0.4$ to 1.4 $M_\odot$, models of PMS evolution that do not include internal magnetic fields ("standard" models) underestimate the actual stellar mass by about 30%.

2) Track masses estimated using an HRD with magnetic evolutionary tracks reproduce the actual masses much more closely. The average difference between the dynamical mass and track mass is $0.01 \pm 0.02$ $M_\odot$.

3) Given this agreement between the dynamical and magnetic track masses, results inferred for both masses and ages using the magnetic tracks are likely more reliable than those derived from the standard tracks.

4) Ages of the stars in the Taurus SFR evaluated using isochrones of the magnetic models are older, $\sim 3$ to 10 MY, than the 1-3 MY range inferred using the standard model. The broad age range is principally the result of uncertainties in the stellar luminosities.

5) The older ages increase the timescale for the formation of giant gas planets and their possible migration to orbits very near the host stars where they appear as the hot Jupiters. The longer timescales also favor the chemical evolution of the massive disks in which the planets form.

We believe that our results are sufficiently secure to recommend the use of magnetic models of stellar evolution to estimate the masses of low-mass PMS stars when dynamical masses are not available. To improve on this work, two advances are necessary. The sample with which worked has very few stars with $M_{dyn} < 0.4$ $M_\odot$. Because there are significant differences among the models at the lowest masses and ages less than a few MY ([Simon & Toraskar 2017](#)), it is necessary to increase the sample at the lowest masses. Finally, a consequence of internal magnetic fields is an increase in the photospheric radius over that predicted by the non-magnetic models. The acid test will come from O/IR interferometric measurements of the diameters of the T Tauris.

*Facility:* GAIA, ALMA

*Software:* CASA, [McMullin et al. 2007](#); GILDAS, [Pety 2005](#); DiskFit, [Piétu et al. 2007](#)

We thank the referee for extremely prompt and helpful reports that resulted in important improvements of our paper. MS thanks D. Pourbaix for much advice about GAIA data, C. Bender for saving his sanity in the color-coding of Fig. 1 and J. Schlieder for help with an earlier form of the figure. S. Brittain kindly provided advice concerning MWC 480's stellar parameters. This work was supported in part by the "Programme National de Physique Stellaire" (PNPS) of CNRS/INSU co-funded by CEA and CNES. Our work has made use of data from the European Space Agency (ESA) mission Gaia (https://www.cosmos.esa.int/gaia) processed by the Gaia Data Processing and Analysis Consortium (DPAC https://www.cosmos.esa.int/web/gaia/dpac/consortium). Funding for the DPAC has been provided by national institutions, in particular the institutions participating in the Gaia Multilateral Agreement. Our work has also used ALMA archival data as specified in the text. ALMA is a partnership of ESO (representing its member states), NSF (USA) and NINS (Japan), together with NRC (Canada) and NSC and ASIAA (Taiwan), in cooperation with the Republic of Chile. The Joint ALMA Observatory is operated by ESO, AUI/NRAO and NAOJ.

**Table 1**. Masses of Stars in the Taurus Star Forming Region

| Name | Gal. l | Gal. b | $M_{pub}$ | $D_{pub}$ | D(DR2) | $M_{dyn}$ | Ref. |
|---|---|---|---|---|---|---|---|
| | deg | deg | $M_\odot$ | pc | pc | $M_\odot$ | |
| FM Tau | 168.21 | -16.33 | $0.36 \pm 0.01$ | 131 | $131.4 \pm 0.8$ | $0.36 \pm 0.02$ | 1 |
| CW Tau | 168.24 | -16.34 | $0.62 \pm 0.02$ | 131 | $131.9 \pm 0.7$ | $0.64 \pm 0.01$ | 1,2 |
| FP Tau | 169.38 | -17.24 | $0.37 \pm 0.02$ | 131 | $128.0 \pm 0.9$ | $0.36 \pm 0.02$ | 1 |
| CX Tau | 169.36 | -17.22 | $0.37 \pm 0.02$ | 131 | $127.5 \pm 0.7$ | $0.38 \pm 0.02$ | 1,2 |
| CY Tau | 168.64 | -15.71 | $0.31 \pm 0.02$ | 131 | $128.4 \pm 0.7$ | $0.30 \pm 0.02$ | 1 |
| BP Tau | 168.34 | -14.91 | $1.32^{+0.20}_{-0.13}$ | 140 | $128.6 \pm 1.0$ | $1.10 \pm 0.04$ | 3,2 |
| J04202144+2813491 | 169.17 | -15.34 | $0.27 \pm 0.01$ | 140 | N/A | | 2 |
| DE Tau | 169.65 | -15.30 | | | $126.9 \pm 1.1$ | $0.41 \pm 0.03$ | 2 |
| FS Tau B | 179.40 | -15.94 | $0.74 \pm 0.01$ | 140 | N/A | | 2 |
| J04230776+2805573 | 169.70 | -14.98 | | | $138.6^{+3.4}_{-3.2}$ | $0.52 \pm 0.04$ | 2 |
| IP Tau | 170.67 | -15.29 | $0.95 \pm 0.05$ | 131 | $130.1 \pm 0.7$ | $0.94 \pm 0.05$ | 1 |
| FV Tau A | 171.81 | -15.70 | $2.30 \pm 0.17$ | 140 | N/A | | 1 |
| IQ Tau | 172.26 | -15.20 | $0.79 \pm 0.02$ | 140 | $130.8 \pm 1.1$ | $0.74 \pm 0.02$ | 4 |
| FX Tau A | 173.67 | -16.19 | $1.70 \pm 0.18$ | 140 | N/A | | 1 |
| DK Tau A | 172.47 | -15.11 | $0.60 \pm 0.14$ | 140 | $128.1 \pm 1.0$ | $0.55 \pm 0.13$ | 1 |
| DK Tau B | 172.47 | -15.11 | $1.30 \pm 0.30$ | 140 | $128.1 \pm 1.0$ | $1.19 \pm 0.27$ | 1 |
| HK Tau A | 173.91 | -15.98 | $0.58 \pm 0.05$ | 140 | $132.8 \pm 1.6$ | $0.53 \pm 0.03$ | 1,2 |
| HK Tau B | 173.91 | -15.98 | $1.00 \pm 0.03$ | 140 | $132.8 \pm 1.6$ | $0.89 \pm 0.04$ | 1,2 |
| V710 Tau A | 178.81 | -19.83 | $0.66 \pm 0.06$ | 140 | $142.4 \pm 2.2$ | $0.67 \pm 0.06$ | 1 |
| Haro 6-13 | 173.91 | -15.86 | $1.00 \pm 0.15$ | 140 | $130.1 \pm 3.1$ | $0.93 \pm 0.14$ | 4 |
| GK Tau | 174.22 | -15.71 | $0.79 \pm 0.07$ | 140 | $128.8 \pm 0.7$ | $0.73 \pm 0.06$ | 1 |
| DL Tau | 173.45 | -15.06 | $1.05 \pm 0.02$ | 161 | $158.6 \pm 1.2$ | $1.04 \pm 0.02$ | 1 |
| HN Tau A | 179.48 | -19.83 | $1.57 \pm 0.15$ | 140 | $136.1 \pm 3.1$ | $1.53 \pm 0.15$ | 1 |
| DM Tau | 179.26 | -19.61 | $0.53 \pm 0.02$ | 140 | $144.5 \pm 1.1$ | $0.55 \pm 0.02$ | 4 |
| CI Tau | 175.46 | -16.64 | $0.92 \pm 0.02$ | 161 | $158.0 \pm 1.2$ | $0.90 \pm 0.02$ | 1 |
| IT Tau B | 172.79 | -14.44 | $0.50 \pm 0.08$ | 140 | $161.3 \pm 2.0$ | $0.58 \pm 0.09$ | 1 |
| AA Tau | 174.32 | -15.40 | $0.82 \pm 0.02$ | 140 | $136.7 \pm 2.5$ | $0.84 \pm 0.04$ | 1,2 |
| HO Tau | 175.93 | -16.57 | $0.37 \pm 0.03$ | 140 | $161.3 \pm 1.2$ | $0.43 \pm 0.03$ | 1 |
| DN Tau | 174.59 | -15.45 | $0.95 \pm 0.16$ | 140 | $127.8 \pm 0.9$ | $0.87 \pm 0.15$ | 4 |
| HBC 411 B | 174.67 | -15.45 | $2.05 \pm 0.20$ | 140 | $124.8 \pm 2.2$ | $1.83 \pm 0.18$ | 1 |
| HV Tau C | 173.53 | -13.67 | $1.43 \pm 0.03$ | 140 | $133.5 \pm 2.0$ | $1.33 \pm 0.04$ | 1, 2 |
| LkCa 15 | 176.69 | -15.98 | $1.01 \pm 0.03$ | 140 | $158.2 \pm 1.2$ | $1.14 \pm 0.03$ | 4 |
| Haro 6-33 | 174.17 | -13.29 | $0.5 \pm 0.1$ | 140 | $160.1^{+10.8}_{-9.5}$ | $0.6 \pm 0.1$ | 4 |
| GO Tau | 174.85 | -13.43 | $0.48 \pm 0.01$ | 140 | $144.0 \pm 1.0$ | $0.45 \pm 0.01$ | 4 |
| DS Tau | 172.38 | -10.03 | $0.73 \pm 0.02$ | 140 | $158.4 \pm 1.1$ | $0.83 \pm 0.02$ | 1 |
| GM Aur | 172.57 | -8.19 | $1.14 \pm 0.02$ | 159 | $159.0 \pm 2.1$ | $1.14 \pm 0.02$ | 1 |
| MWC 480 | 173.46 | -7.90 | $1.83 \pm 0.05$ | 140 | $161.1 \pm 2.0$ | $2.11 \pm 0.06$ | 5 |
| CIDA-9 A | 177.85 | -9.33 | $1.08 \pm 0.20$ | 140 | $171.0 \pm 2.7$ | $1.32 \pm 0.24$ | 1 |
| CQ Tau | 182.41 | -4.04 | | | $162.4 \pm 2.2$ | $1.51 \pm 0.07$ | 2 |

References: 1) Simon et al. (2017), 2) This paper, 3) Dutrey et al. (2003), 4) Guilloteau et al. (2014), 5) Schaefer et al. (2009)

**Table 2.** Masses of Stars in the Ophiuchus Star Forming Region

| Name | Gal. l | Gal. b | $M_{pub}$ | $D_{pub}$ | D(DR2) | $M_{dyn}$ | Ref. |
|------|--------|--------|-----------|-----------|--------|-----------|------|
|      | deg    | deg    | $M_\odot$ | pc        | pc     | $M_\odot$ |      |
| GSS 26 | 353.07 | +16.99 | $1.51 \pm 0.02$ | 119 | N/A | | 1 |
| GSS 39 | 353.13 | +16.87 | $0.47 \pm 0.01$ | 119 | $118.5^{+16.7}_{-13.1}$ | $0.47 \pm 0.04$ | 1 |
| YLW 16C | 353.03 | +16.57 | $1.80 \pm 0.10$ | 119 | N/A | | 1 |
| ROX 25 | 353.02 | +16.53 | $1.10 \pm 0.07$ | 119 | $137.5 \pm 3.3$ | $1.27 \pm 0.07$ | 1 |
| YLW 58 | 353.19 | +16.45 | $0.09 \pm 0.01$ | 119 | $136.6^{+2.7}_{-2.5}$ | $0.10 \pm 0.01$ | 1 |
| Flying Saucer | 353.25 | +16.25 | $0.58 \pm 0.01$ | 120 | N/A | | 1 |

Reference: 1) Simon et al. (2017)

**Table 3.** PMS Stars in Taurus SFR: $M_{dyn}$, $L/L_\odot$, and $T_{eff}$

| Name | $M_{dyn}$ | Lu18 SpTy | And13 L | And13 $T_{eff}$ | He14 L | He14 $T_{eff}$ |
|------|-----------|-----------|---------|-----------------|--------|----------------|
|      | $M_\odot$ |           | $L_\odot$ | K             | $L_\odot$ | K           |
| FM Tau | $0.36 \pm 0.02$ | M4.5 | $0.40 \pm 0.09$ | $3125 \pm 125$ | N/A | |
| CW Tau | $0.64 \pm 0.01$ | K3 | $2.42 \pm 1.40$ | $4840 \pm 100$ | $0.45 \pm 0.21$ | $4539 \pm 80$ |
| FP Tau | $0.36 \pm 0.02$ | M2.6 | $0.32 \pm 0.04$ | $3460 \pm 120$ | $0.17 \pm 0.10$ | $3470 \pm 49$ |
| CX Tau | $0.38 \pm 0.02$ | M2.5 | $0.38 \pm 0.05$ | $3475 \pm 130$ | $0.25 \pm 0.12$ | $3485 \pm 49$ |
| CY Tau | $0.30 \pm 0.02$ | M2.3 | $0.40 \pm 0.08$ | $3500 \pm 150$ | $0.26 \pm 0.12$ | $3500 \pm 49$ |
| BP Tau | $1.10 \pm 0.04$ | M0.5 | $0.81 \pm 0.27$ | $3935 \pm 160$ | $0.40 \pm 0.23$ | $3840 \pm 90$ |
| J04202144+2813491 | $0.27 \pm 0.01$ | M2 | N/A | | | |
| DE Tau | $0.41 \pm 0.03$ | M2.3 | $0.85 \pm 0.15$ | $3550 \pm 150$ | $0.49 \pm 0.28$ | $3515 \pm 49$ |
| J04230776+2805573 | $0.52 \pm 0.04$ | M2 | N/A | | | |
| IP Tau | $0.94 \pm 0.05$ | M0.6 | $0.47 \pm 0.10$ | $3750 \pm 180$ | $0.47 \pm 0.10$ | $3828 \pm 54$ |
| IQ Tau | $0.74 \pm 0.02$ | M1.1 | $0.81 \pm 0.17$ | $3660 \pm 160$ | $0.22 \pm 0.10$ | $3750 \pm 50$ |
| HK Tau A | $0.53 \pm 0.03$ | M1 | $0.55 \pm 0.15$ | $3680 \pm 150$ | N/A | |
| HK Tau B | $0.89 \pm 0.04$ | M2 | $0.03 \pm 0.01$ | $3550 \pm 150$ | N/A | |
| V710 Tau A | $0.69 \pm 0.06$ | M3.3 | $0.57 \pm 0.19$ | $3340 \pm 130$ | $0.38 \pm 0.15$ | $3349 \pm 68$ |
| GK Tau | $0.73 \pm 0.06$ | K6.5 | $1.34 \pm 0.97$ | $4130 \pm 150$ | $0.79 \pm 0.29$ | $4068 \pm 50$ |
| DL Tau | $1.04 \pm 0.02$ | K5.5 | $0.74 \pm 0.25$ | $4325 \pm 200$ | $0.64 \pm 0.22$ | $4163 \pm 50$ |
| HN Tau A | $1.53 \pm 0.15$ | K5 | $0.42 \pm 0.30$ | $4450 \pm 160$ | $0.17 \pm 0.07$ | $4210 \pm 80$ |
| DM Tau | $0.55 \pm 0.02$ | M3 | $0.23 \pm 0.01$ | $3440 \pm 130$ | $0.13 \pm 0.06$ | $3410 \pm 30$ |
| CI Tau | $0.90 \pm 0.02$ | K5.5 | $0.92 \pm 0.25$ | $4325 \pm 200$ | $0.63 \pm 0.20$ | $4163 \pm 50$ |
| AA Tau | $0.84 \pm 0.04$ | M0.6 | $0.84 \pm 0.40$ | $3748 \pm 170$ | $0.43 \pm 0.16$ | $3760 \pm 40$ |
| HO Tau | $0.43 \pm 0.03$ | M3.2 | $0.13 \pm 0.06$ | $3360 \pm 130$ | $0.14 \pm 0.06$ | $3365 \pm 68$ |
| HBC 411 B | $1.83 \pm 0.18$ | M0.5 | $0.20 \pm 0.07$ | $3770 \pm 170$ | N/A | |
| HV Tau C | $1.33 \pm 0.04$ | K6 | $0.03 \pm 0.02$ | $4200 \pm 200$ | N/A | |
| LkCa 15 | $1.14 \pm 0.03$ | K5.5 | $0.81 \pm 0.39$ | $4325 \pm 200$ | $0.78 \pm 0.30$ | $4163 \pm 50$ |
| GO Tau | $0.45 \pm 0.01$ | M2.3 | $0.29 \pm 0.09$ | $3500 \pm 100$ | $0.21 \pm 0.12$ | $3515 \pm 48$ |
| DS Tau | $0.83 \pm 0.02$ | M0.4 | $0.76 \pm 0.34$ | $3782 \pm 170$ | $0.25 \pm 0.09$ | $3810 \pm 62$ |
| GM Aur | $1.14 \pm 0.02$ | K6 | $1.23 \pm 0.31$ | $4200 \pm 250$ | $0.63 \pm 0.22$ | $4115 \pm 50$ |
| MWC 480 | $2.11 \pm 0.06$ | A2 | $25.3^{+22.2}_{-11.8}$ | $8329 \pm 400$ | N/A | |
| CQ Tau | $1.51 \pm 0.07$ | F5 | $11.9 \pm 1.5$ | $6800 \pm 200$ | N/A | |

**Table 4.** PMS Stars in Ophiuchus SFR: $M_{dyn}$, $L/L_\odot$, and $T_{eff}$

| Name | $M_{dyn}$ | Ri10, Er11 | | | Na15 | | |
|---|---|---|---|---|---|---|---|
| | | SpTy | L | $T_{eff}$ | SpTy | L | $T_{eff}$ |
| | $M_\odot$ | | $L_\odot$ | K | | $L_\odot$ | K |
| GSS 39 | $0.47 \pm 0.04$ | M0 | $0.78^{+0.21}_{-0.17}$ | $3850 \pm 100$ | M0 | $1.29^{+0.95}_{-0.55}$ | $3850 \pm 100$ |
| ROX 25 | $1.27 \pm 0.07$ | K6 | $0.58 \pm 0.03$ | $4200 \pm 200$ | K7 | $3.71 \pm 0.16$ | $4050 \pm 200$ |
| YLW 58 | $0.10 \pm 0.01$ | M4 | $0.27 \pm 0.01$ | $3200 \pm 100$ | M4.5 | $0.24 \pm 0.02$ | $3125 \pm 75$ |

**Table 5.** $M_{dyn}$ and $M_{hrd}$ in the Taurus SFR

| Name | $M_{dyn}$ | $MA_{hrd}$ | $MB_{hrd}$ | $MC_{hrd}$ |
|---|---|---|---|---|
| | $M_\odot$ | $M_\odot$ | $M_\odot$ | $M_\odot$ |
| FM Tau | $0.36 \pm 0.02$ | $0.18 \pm 0.02$ | | |
| CW Tau | $0.64 \pm 0.02$ | $1.45^{+0.05}_{-0.20}$ | $1.05 \pm 0.07$ | |
| FP Tau | $0.36 \pm 0.02$ | $0.31 \pm 0.05$ | $0.32 \pm 0.08$ | $0.47 \pm 0.04$ |
| CX Tau | $0.38 \pm 0.02$ | $0.32 \pm 0.06$ | $0.33 \pm 0.04$ | $0.47 \pm 0.04$ |
| CY Tau | $0.30 \pm 0.02$ | $0.33 \pm 0.07$ | $0.34 \pm 0.03$ | $0.49 \pm 0.04$ |
| BP Tau | $1.10 \pm 0.04$ | $0.60 \pm 0.12$ | $0.58 \pm 0.09$ | $0.88 \pm 0.10$ |
| DE Tau | $0.41 \pm 0.03$ | $0.38 \pm 0.06$ | $0.34 \pm 0.04$ | $0.45 \pm 0.06$ |
| IP Tau | $0.94 \pm 0.05$ | $0.40 \pm 0.10$ | $0.55 \pm 0.05$ | $0.80 \pm 0.05$ |
| IQ Tau | $0.74 \pm 0.02$ | $0.40 \pm 0.10$ | $0.54 \pm 0.05$ | $0.70 \pm 0.05$ |
| HK Tau A | $0.53 \pm 0.03$ | $0.42^{+0.10}_{-0.05}$ | | |
| V710 Tau A | $0.67 \pm 0.06$ | $0.32 \pm 0.05$ | $0.28 \pm 0.04$ | $0.32 \pm 0.04$ |
| GK Tau | $0.73 \pm 0.06$ | $0.78 \pm 0.10$ | $0.71 \pm 0.10$ | $1.10 \pm 0.08$ |
| DL Tau | $1.04 \pm 0.02$ | $1.00 \pm 0.15$ | $0.87 \pm 0.05$ | $1.15 \pm 0.05$ |
| HN Tau A | $1.53 \pm 0.15$ | $1.05 \pm 0.15$ | | |
| DM Tau | $0.55 \pm 0.02$ | $0.30 \pm 0.02$ | $0.32 \pm 0.02$ | $0.40 \pm 0.02$ |
| CI Tau | $0.90 \pm 0.02$ | $1.00 \pm 0.20$ | $0.80 \pm 0.05$ | $1.15 \pm 0.05$ |
| AA Tau | $0.84 \pm 0.04$ | $0.48 \pm 0.06$ | $0.50 \pm 0.09$ | $0.72 \pm 0.08$ |
| HO Tau | $0.43 \pm 0.03$ | $0.29 \pm 0.090$ | $0.28 \pm 0.04$ | $0.38 \pm 0.08$ |
| HBC 411 B | $1.83 \pm 0.18$ | $0.58 \pm 0.10$ | | |
| LkCa 15 | $1.14 \pm 0.04$ | $1.00 \pm 0.20$ | $0.80 \pm 0.04$ | $1.20 \pm 0.05$ |
| GO Tau | $0.45 \pm 0.01$ | $0.35 \pm 0.05$ | $0.35 \pm 0.03$ | $0.50 \pm 0.05$ |
| DS Tau | $0.83 \pm 0.02$ | $0.50 \pm 0.10$ | $0.58 \pm 0.10$ | $0.80 \pm 0.05$ |
| GM Aur | $1.14 \pm 0.03$ | $0.80^{+0.50}_{-0.15}$ | $0.78 \pm 0.08$ | $1.15 \pm 0.08$ |
| MWC 480 | $2.11 \pm 0.06$ | $2.05^{+0.45}_{-0.15}$ | | |
| CQ Tau | $1.51 \pm 0.07$ | $1.75 \pm 0.03$ | | |

$MA_{hrd}$ and $MB_{hrd}$ using Fig. 3, $MC_{hrd}$ using Fig.8A

**Table 6.** $M_{dyn}$ and $M_{hrd}$ in the Ophiuchus SFR

| Name | $M_{dyn}$ | Ri10 and Er11 | Na15 |
|---|---|---|---|
| | | $MD_{hrd}$ | $ME_{hrd}$ |
| | $M_\odot$ | $M_\odot$ | $M_\odot$ |
| GSS 39 | $0.47 \pm 0.04$ | $0.78 \pm 0.20$ | |
| ROX 25 | $1,27 \pm 0.07$ | $1.15 \pm 0.10$ | |
| YLW 58 | $0.10 \pm 0.01$ | $0.25 \pm 0.08$ | $0.21 \pm 0.08$ |

$MD_{hrd}$ and $ME_{hrd}$ from Fig.8B

**Table 1**. Analyses of ALMA Archival Data

| ID | D(pc) | $V_{lsr}$(km/s) | M(M$_\odot$) | CO Results | | | Continuum Results | | |
| --- | --- | --- | --- | --- | --- | --- | --- | --- | --- |
| | | | | PA | i(CO) | R$_{out}$ | PA | i | R$_{out}$ |
| AA Tau | 136.7 ± 2.5 | 6.50 ± 0.02 | 0.838 ± 0.036 | 3.0 ± 0.2 | −58.3 ± 0.3 | 286 ± 4 | 4.4 ± 0.3 | −37.2 ± 0.1 | 125 ± 1 |
| BP Tau | 128.6 ± 1.0 | 6.73 ± 0.01 | 1.102 ± 0.042 | 239.2 ± 0.4 | 31.0 ± 1.0 | 117 ± 9 | 87.1 ± 0.4 | 45.9 ± 0.2 | 120 ± 10 |
| CQ Tau | 162.4 ± 2.2 | 6.21 ± 0.01 | 1.510 ± 0.068 | 323.9 ± 0.4 | 35.1 ± 1.3 | 146 ± 3 | 299.5 ± 1.1 | 38.0 ± 4.8 | 83 ± 1 |
| CW Tau | 131.9 ± 0.7 | 6.44 ± 0.01 | 0.637 ± 0.013 | 332.5 ± 0.2 | 60.2 ± 5.7 | 351 ± 45 | 333.0 ± 0.6 | 56.2 ± 0.4 | 89 ± 2 |
| CX Tau | 127.5 ± 0.7 | 8.40 ± 0.01 | 0.380 ± 0.020 | 339.0 ± 0.5 | 62.8 ± 1.3 | 151 ± 30 | 154.4 ± 5.0 | 50.7 ± 1.5 | 28 ± 5 |
| DE Tau | 126.9 ± 1.1 | 5.92 ± 0.00 | 0.407 ± 0.030 | 210.0 ± 0.3 | 25.2 ± 0.4 | 213 ± 1 | 214.1 ± 1.8 | 34.1 ± 1.0 | 23 ± 1 |
| FS Tau B | 140.0 ± 0.8 | 7.45 ± 0.03 | 0.739 ± 0.013 | 52.9 ± 1.1 | 80.2 ± 0.4 | 739 ± 18 | 54.7 ± 0.8 | 75.4 ± 0.3 | 137 ± 1 |
| HK Tau A | 133.3 ± 1.6 | 6.10 ± 0.02 | 0.528 ± 0.027 | 88.8 ± 0.6 | 61.8 ± 1.8 | 101 ± 5 | 87.5 ± 4.8 | 63.1 ± 1.5 | 104 ± 15 |
| HK Tau B | 133.3 ± 1.6 | 6.52 ± 0.07 | 0.888 ± 0.040 | 131.7 ± 1.4 | 84.9 ± 1.2 | 115 ± 6 | 131.0 ± 1.8 | 84.7 ± 1.4 | 85 ± 5 |
| HV Tau C | 134.0 ± 2.0 | 6.34 ± 0.01 | 1.332 ± 0.035 | 195.5 ± 0.8 | 86.0 ± 6.8 | 861 ± 3 | 198.7 ± 1.6 | 89.2 ± 70.5 | 93 ± 16 |
| J04202144 +2813491 | default: 140 | 7.38 ± 0.01 | 0.272 ± 0.009 | 73.6 ± 0.2 | −90.6 ± 0.3 | 609 ± 18 | 73.9 ± 1.5 | −90.2 ± 11.2 | 285 ± 14 |
| J04230776 +2805573 | 138.6$^{+3.4}_{-3.2}$ | 6.22 ± 0.02 | 0.524 ± 0.041 | 219.7 ± 10.2 | −61.7 ± 1.6 | 166 ± 6 | 219.7 ± 4.0 | −70.7 ± 2.6 | 78 ± 4 |

## APPENDIX

Table 1 of the Appendix presents the results of our analysis of the ALMA archival data. Columns 1 to 4 list the star's name, Gaia DR2 distance, $V_{lsr}$ measured in the CO lines, and the fitted dynamical mass. Columns 5-7 give the fitted disk parameters as observed in CO, the position angle of the disk rotation angle, inclination, and outer radius. Columns 8-10 provide the disk parameters derived from analysis of the disk continuum emission. Images of the data, model and residuals are presented in Figs. 1 to 11. The cross in each panel indicates the projected major and minor axis of the outer disk radius.

*Project 2013.1.01070.S and 2015.1.010017.S*— The ALMA projects 2013.1.01070.S and 2015.1.010017.S (P.I. R.Loomis) observed HCN, HCO$^+$ and CO, $^{13}$CO and CN in AA Tau respectively. We used here only the $^{13}$CO data for the stellar mass measurement.

The data set provide an angular resolution of $0.25 \times 0.2''$ and spectral resolution $0.22$ km.$^{-1}$. The derived inclination ($58.1 \pm 0.2°$) and orientation (PA $3 \pm 0.3°$ for the rotation axis) are in good agreement with Loomis et al. (2017)'s analysis. The best fit exhibit small systematic residuals (see Fig. 1) that correspond to the warp inside $\sim 50$ AU (Loomis et al., 2017) but these small deviations do not affect the derived kinematics and stellar mass, since the data is largely dominated by the outer parts.

*Project 2011.0.00320*— The ALMA project 2011.0.00320 (P.I. E.Chapillon) observed the CO and $^{13}$CO J=3-2 lines towards 3 sources, BP Tau, CQ Tau and MWC 758. The spectral resolution is about 0.11 km.s$^{-1}$. The initial data was affected by a velocity shift. Re-calibrating the whole data set using a more recent version of CASA suppressed this issue. The angular resolution is about $0.7 \times 0.5''$ at PA $20°$. This is limited compared to the derived disk sizes (outer radii around 150 au, i.e. less than $2''$ in size). This limits the precision of the disk inclination measurement, especially from the continuum emission.

MWC 758 has a very asymmetric continuum emission dominated by emission from dense spots in a spiral pattern (Boehler et al. 2018) which makes it impossible to derive an inclination. Furthermore, the CO emission suggests a relatively low disk inclination, so no stellar mass can be confidently derived from this data set.

For BP Tau, the signal to noise ratio in $^{13}$CO 3-2 is not sufficient to perform a complete fit, but the data confirms that the systemic velocity derived from $^{12}$CO is accurate enough for our purpose. Contamination by the molecular cloud, which was quite important in the earlier CO J=2-1 data by Dutrey et al. (2003) is very limited here (see Fig. 2). The inclination derived from the continuum differs from that of the CO disk. This could be due to sub-structures similar to those seen in MWC 758. Given the small disk size, we do not consider this discrepancy as sufficiently important to cast a doubt on the stellar mass derivation. This new data confirms (with higher accuracy because of better signal to noise and lower contamination) the overall disk properties derived by Dutrey et al. (2003) but yield a lower temperature, around 20 K at 100 au.

For CQ Tau (Fig. 3 Appendix), $^{13}$CO and $^{12}$CO data give consistent results, but the $^{12}$CO residuals indicate significant substructures, perhaps due to a disk warp. As for BP Tau, the small difference in orientation and inclination between the continuum and line emission is not significant for our purpose but could indicate incompletely resolved continuum sub-structures.

*Project 2013.1.00426* — The ALMA project 2013.1.00426 (P.I. Y.Boehler) observed the CO J=2-1 and J=3-2 lines towards a few disks with low mm continuum flux. An analysis of CX Tau and FM Tau was already reported in Simon et al. (2017). The CO J=3-2 data was re-analyzed here with the proper distance and represented in the same format as other sources for consistency.

We also re-analyzed the CO J=3-2 data of DE Tau, for which an angular resolution of $0.28 \times 0.19''$ at PA 25° was obtained. The spectral resolution is about 0.21 km.s$^{-1}$. There is significant contamination by the molecular cloud and a non Keplerian component. However, almost half of the disk projects at velocities beyond those of the cloud, so that a proper analysis remains possible through adequate masking of the contaminated velocity channels, especially because the disk is quite well spatially resolved (see Fig.5).

*Project 2016.1.00460* — The ALMA project 2016.1.00460 (P.I. F.Ménard) observed the CO J=3-2 line of several disks which were known or suspected to be close to edge-on. The CO J=3-2 line was observed with a velocity sampling is 0.21 km/s, leading to an effective resolution about 0.4 km/s. The continuum was covered with 3 additional wide bands.

Data was calibrated using the standard pipeline in the CASA software package and then exported through UVFITS format to the GILDAS package for imaging and data analysis. The typical synthesized beam is $0.5 \times 0.35''$. The line sensitivity is about 20 mJy/beam, or about 1.2 K. No self-calibration was performed, but the data is of sufficient quality for our analysis.

The initial sample included sources which had previously been observed in other spectral lines, such as HV Tau, HK Tau and IRAS04302+2247 (the Butterfly star), as well as stars for which no dynamical mass estimate were available. Specific comments on individual sources in this data set are given below.

*IRAS04302+2247* — (the Butterfly star): there is strong contamination by the molecular cloud, and a clear evidence for a relatively faint outflow perpendicular to the jet, with high velocity bullets. As a result, it is not possible to derive an accurate sytemic velocity, and a fortiori, the rotation velocity. Stellar masses between 1.3 and 1.7 M$_\odot$ remain possible from this data set.

*HK Tau* — The same procedure (alternate subtraction of the best fit model of A and B) as in Simon et al. (2017) was used. These new results confirm the previous measurements, with a somewhat higher accuracy given the improved angular and spectral resolution (see Figs. 6,7).

*HV Tau* — Only HV Tau C is detected, as in our previous study of CN and CO isotopologues (Guilloteau et al. 2014). Surprisingly, CO is only detected near the disk mid-plane. A strong North/South asymmetry is also apparent (see Fig. 8).

*IRAS I04158+2805* — This is a strange object, perhaps a binary, with star(s) of spectral type M5 (Andrews et al. 2008). The circumstellar environment is very unusual, with an asymmetric mm dust ring around the object. These observations resolve the dust cavity and identify an inner disk for the first time. The striking cross visible in CO is actually along the edge of a reflection nebula cavity (Glauser et al. 2008) so its kinematics are not necessarily due to Keplerian rotation. Rotation is visible in the inner regions, but the angular resolution is insufficient to derive a precise dynamical mass. An odd result is the total absence of CO in the circumbinary ring. Hence, no dynamical mass determination was performed on this object.

*2MASS J04220+2650* — This is also known as FS Tau B (see Fig. 9). Kirchschlager et al. (2016) indicate an inclination of 80° from scattered light modeling, in excellent agreement with our derivation. The disk structure is however complex.

*SSTtau 042021+281350* — (2MASS J04202144+2813491) is even more edge-on than the Butterfly star. The large aspect ratio of the mm emission is a clear indication of dust settling in a almost perfectly edge-on disk. However, contrary to HV Tau C, CO is at very high heights above the disk (see Fig.10).

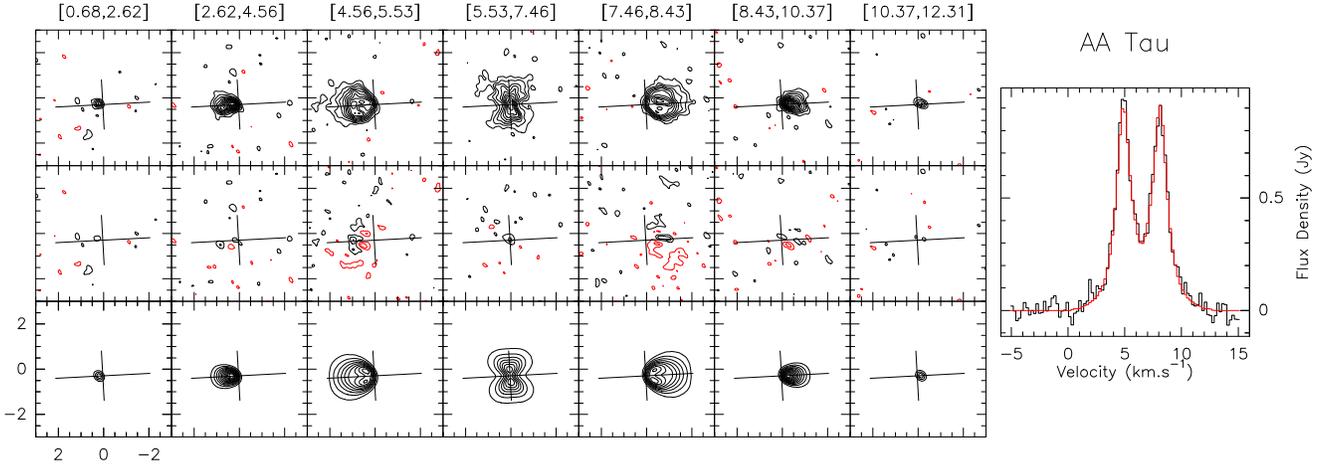

**Figure 1.** $^{13}$CO J=2-1 in AA Tau. Left: Integrated intensity maps in selected velocity ranges. Top row: data; bottom row: model; middle row: residuals. The cross in each panel i ndicates the projected major and minor axis of the outer disk radius. Contours are 3 $\sigma$, with negative contours in red. Coordinates are offsets (in arcsec) from the phase tracking center. The velocity ranges are indicated on top of each column. Right: integrated spectra over the disk size (observed in black; model in red). The grey shaded region indicates the velocity range ignored while fitting the data.

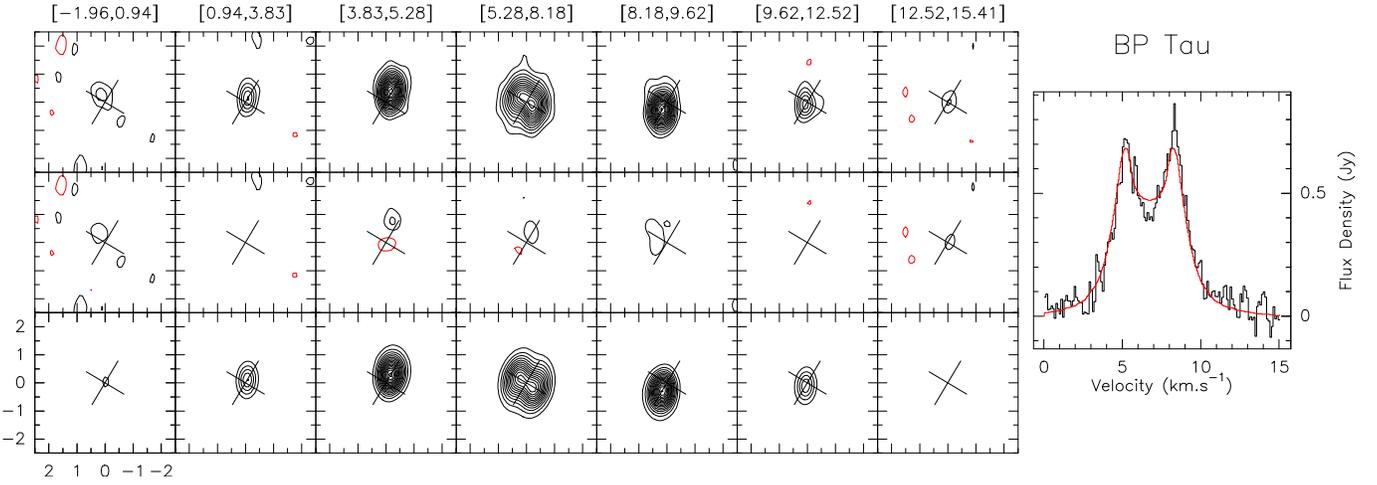

**Figure 2.** Same as Fig. 1 but for BP Tau, CO J = 3-2 .

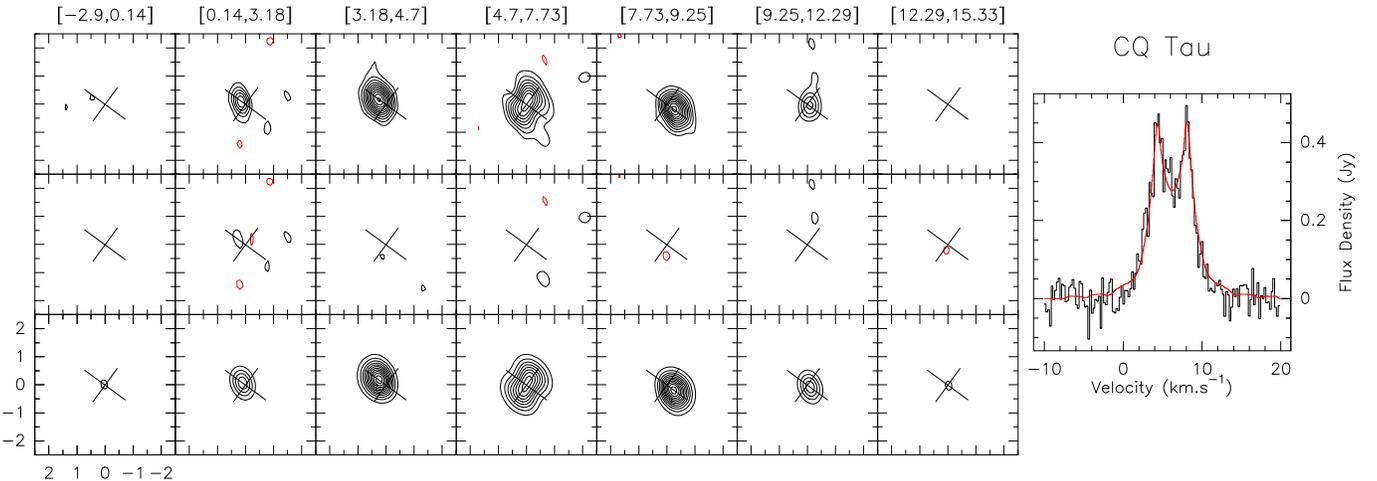

**Figure 3.** Same as Fig. 1 but for CQ Tau, CO J = 3-2.

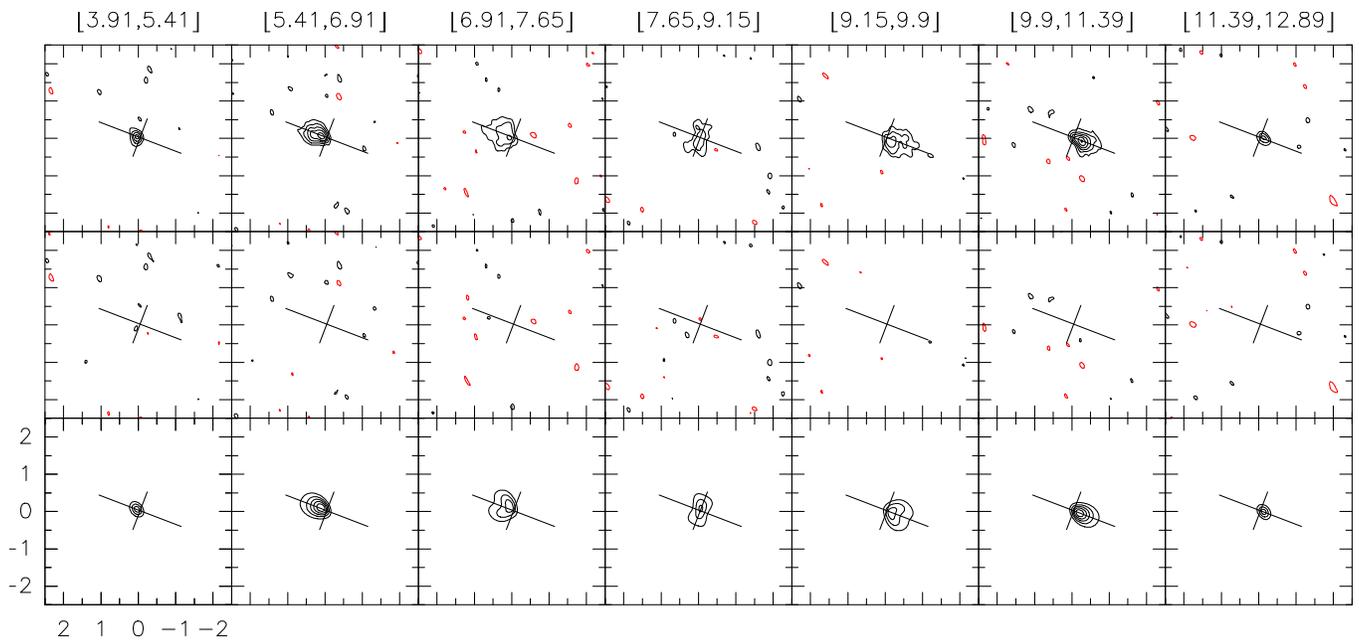

**Figure 4**. Same as Fig. 1 but for CX Tau, CO J = 3-2.

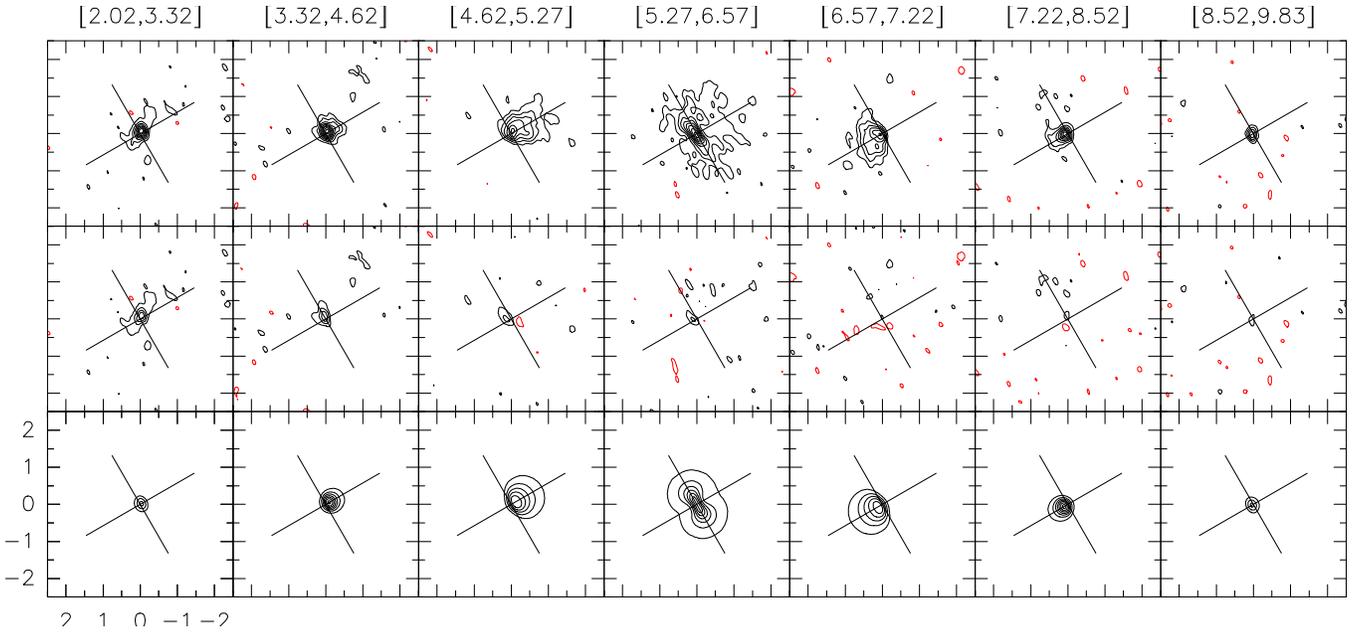

**Figure 5**. Same as Fig. 1 but for DE Tau, CO J = 3-2. The shaded region represents the velocity region excluded from the fit.

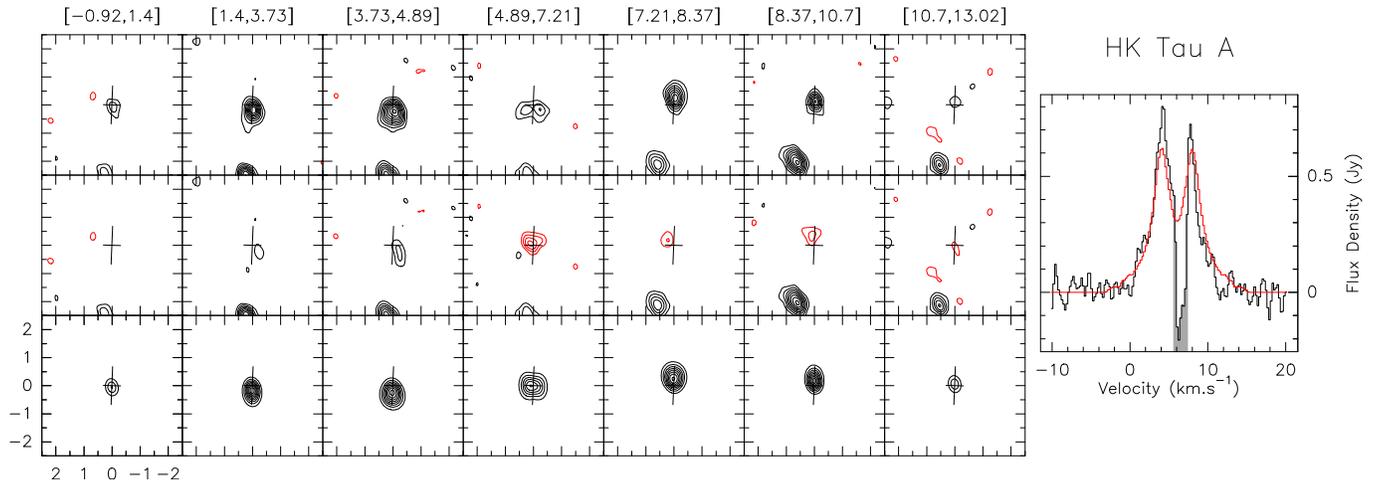

**Figure 6.** Same as Fig. 1 but for HK Tau A, CO J = 3-2.

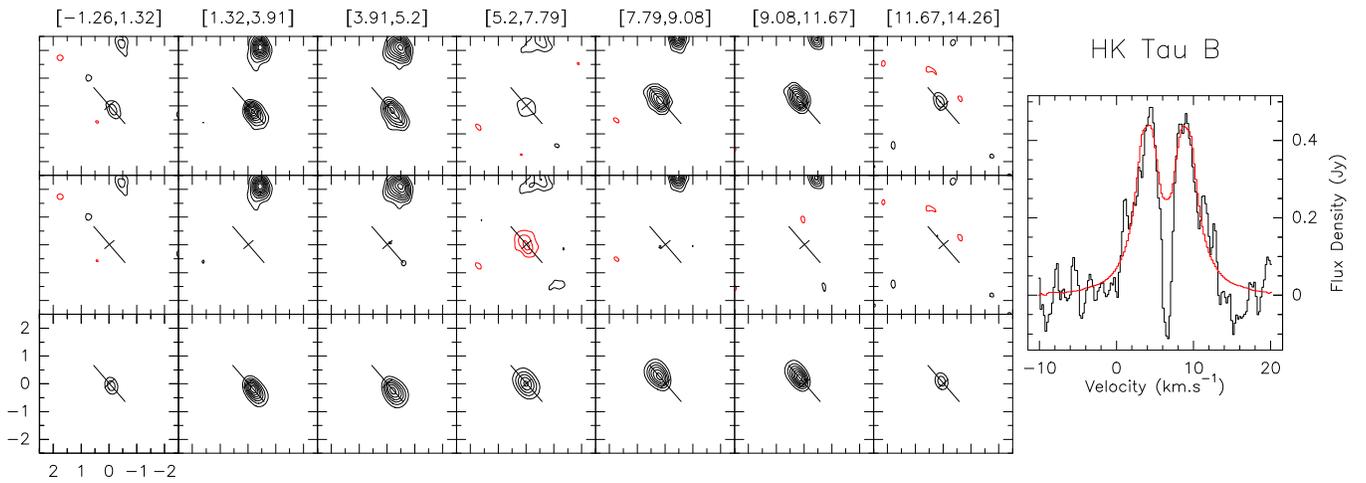

**Figure 7.** Same as Fig. 1 but for HK Tau B, CO J = 3-2.

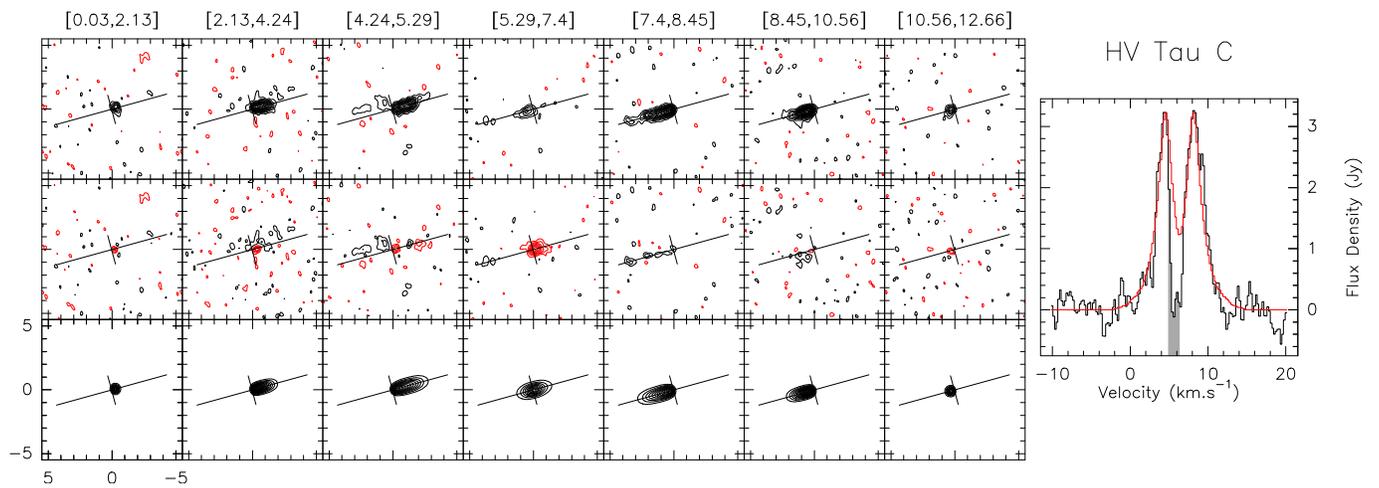

**Figure 8.** Same as Fig. 1 but for HV Tau C, CO J = 3-2.

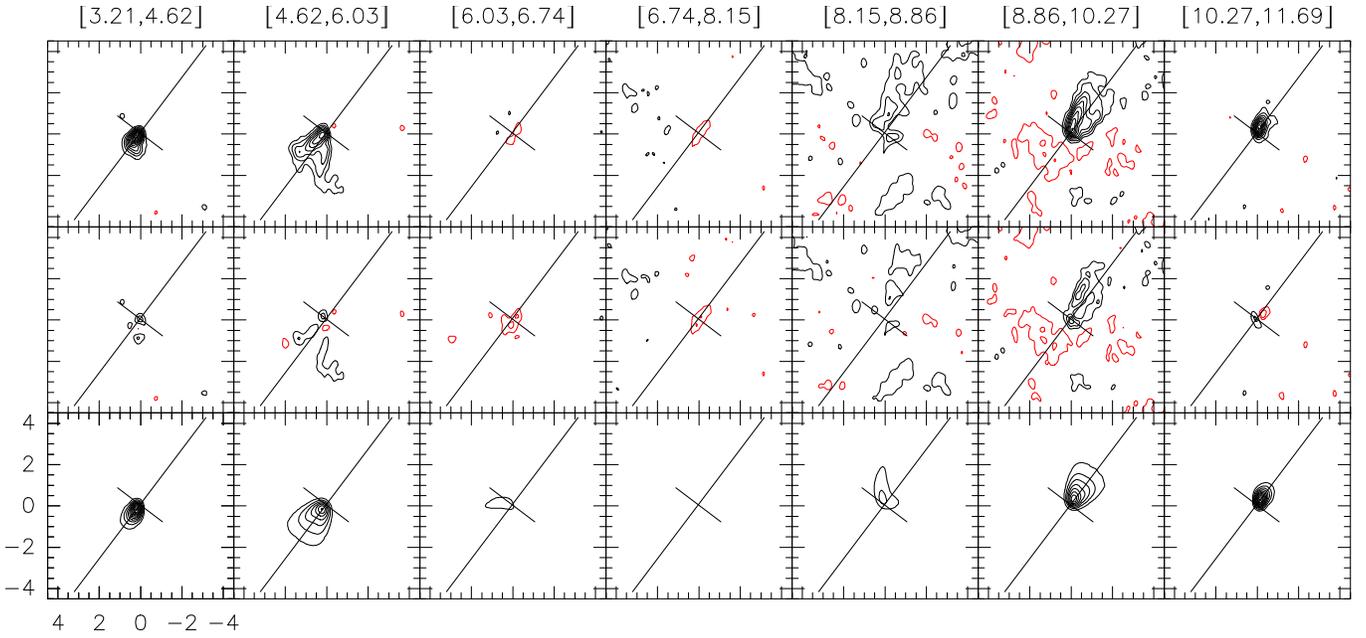

**Figure 9**. As in Fig. 1 but for CO J=3-2 in FS Tau B. The two velocity ranges [6.03,6.74] and [6.74,8.15] are heavily contaminated by the surrounding cloud, and have much higher noise than the others, so the difference appear small in term of Signal to Noise.

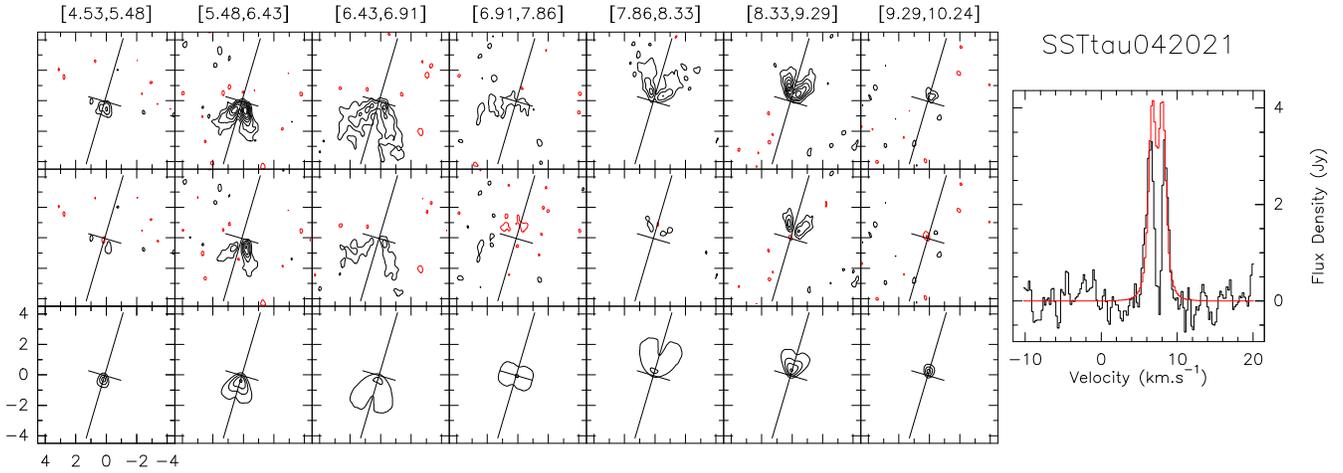

**Figure 10**. As in Fig. 1 but for CO J=3-2 in 2MASS J04202144+2813491 (SSTtau042021) .

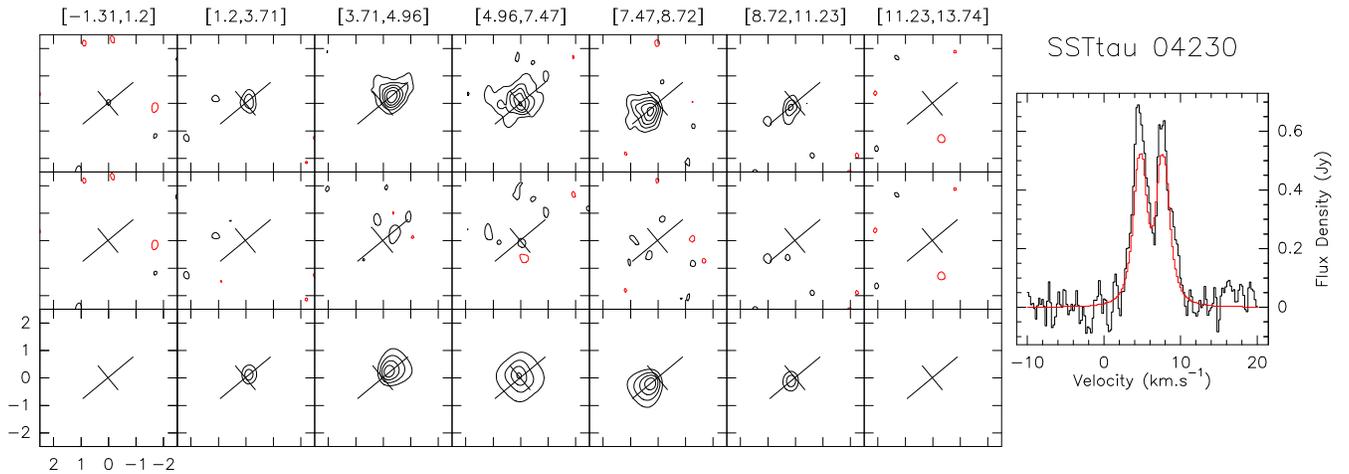

**Figure 11**. As in Fig. 1 but for CO J=3-2 in 2MASS J04230778+2805573 (SSTtau042308.

*SSTtau 042308+280557*— (IRAS 04200+2759) is much less extended than the other sources. Our analysis indicates a rather moderate inclination (see Fig.11).

## REFERENCES

Andrews, Sean, M.; Liu, Michael, C.; Williams, Jonathan, P.; Allers, K. N. 2008, ApJ, 685, 1039

Andrews, Sean M.; Rosenfeld, Katherine A.; Kraus, Adam L.; Wilner, David J. 2013, ApJ, 771, 129 (And13)

Bailer-Jones, C. A. L.; Rybizki, J.; Fouesneau, M.; Mantelet, G.; Andrae, R. 2018, AJ, 156, 59B

Baraffe, I.; Homeier, D;, Allard, F. & Chabrier, G., 2015, A&A, 577, 42

Bell, Cameron P. M.; Naylor, Tim; Mayne, N. J.; et al., 2013, MNRAS, 434, 806

Boehler, Y., Ricci, L., Weaver, E. et al., 2018, ApJ, 853. 162

Choi, J., Dotter, A., Conroy, C., Cantiello, M. Paxton, B., Johnson, B. 2016, ApJ, 823, 102

David, Trevor J.; Hillenbrand, Lynne A.; Gillen, Edward, et al., 2019, ApJ, 872, 161

Donati, J. -F.; Jardine, M. M.; Gregory, S. G. et al. 2008, MNRAS, 386, 1234D

Dutrey, A.;Guilloteau, S.; Simon, M. 2003, A&A, 402, 1003

Erickson, K.L, Wilking, B.A., Meyer, M.R. 2011, , 142, 140 (Er11)

Favre, Cécile, Cleeves, L. Ilsedore; Bergin,Edwin; et al. 2013, ApJ, 776L, 38F

Feiden, G. A. 2016, A&A, 593, 99

Flagg, L., Johns-Krull, C. M., Nofi, L., et al. 2019, ApJL, 878, L37

Glauser, A., Menard, F. Pinte, C. et al. 2008, A&A, 485, 531

Gudel, M.; Briggs, K. R.; Arzner, K.; et al. 2007 A&A, 468, 353

Guilloteau, S.; Simon, M; Piétu, V. et al. 2014, A&A, 567, 117

Gully-Santiago, Michael A.; Herczeg, Gregory J.; Czekala, Ian; et al. 2017 ApJ, 836, 200

Hartigan, Patrick; Edwards, Suzan; Ghandour, Louma 1995 ApJ, 452, 736

Herczeg, Gregory J.; Hillenbrand, Lynne A. 2014, ApJ, 786, 97H (He14)

Hillenbrand, Lynne A.; White, Russel J. 2004, ApJ, 604, 741

Hillenbrand, Lynne A. 2008, Physica Scripta, 130, 4024

Jeffries, R.D., 2017, Mem. della Soc. Ast. Ital., 88, 637

Jeffries, R.D., Jackson, R.J., Franciosini, E. et al 2017 MNRAS, 464, 1456

Johns-Krull, Christopher M.; McLane, Jacob N.; Prato, L. et al. 2016 ApJ, 826, 206J

Kenyon, S.J.; Dobrzyckaz, D.; Hartmann, L. 1994, AJ, 108,1872K

Kirchschlager, F., Wolf, S., Madlener, D. 2016, MNRAS, 462, 858

Lissauer, Jack J.; Hubickyj, Olenka; D'Angelo, Gennaro; Bodenheimer, P. 2009, Icarus, 199, 388

Loomis, R.A. 2017, Oberg, K.I, Andrews, S.M., & Meredith A. ApJ, 840, 23L

Luhman, K.L. 2018 AJ, 156, 271L (Lu18)

MacDonald, James; Mullan, D. J. 2017, ApJ, 850, 58

Malo, Lison; Doyon, Ren; Feiden, Gregory A. et al., 2014, ApJ, 792, 37

Mayne, N.J., & Naylor, Tim, 2008, MNRAS, 386, 261

McMullin, J.P.; Waters, B.; Schiebel, D. et al. ASP Conference Series 2007, Vol. 376, 127

Meyer, M.,; Backman, D., Weinberger, A., &Wyatt, M. in *Protostars and Planets V*, B. Reipurth, Jewitt, D., & Keil, K. (Tucson: U. of AZ Press) p. 538 XS

Muzerolle, James; Hartmann, Lee; Calvet, Nuria 1998 AJ, 116, 455

Najita, Joan R.; Andrews, Sean M.; Muzerolle, James 2015, MNRAS, 450, 3559 (Na15)

Naylor, Tim 2009, MNRAS, 399, 432

Nguyen, Duy Cuong; Brandeker, Alexis; van Kerkwijk, Marten H.; Jayawardhana, Ray 2012 , ApJ, 745, 119

Papaloizou, J. C. B. & Nelson, R. P., 2005, A&A, 433,247

Paxton, B.; Marchant,P.; Schwab, J. et al. 2015, ApJS, 220, 15

Pecaut, Mark J.; Mamajek, Eric E. 2013, ApJS, 208, 9P

Pety, J., Semaine de l'Astrophysique Francaise, 2005, 721

Piétu, V., Dutrey, A., Guilloteau, S. 2007, A&A, 467, 163

Prato, L.; Ruíz-Rodríguez, Dary; Wasserman, L.H. 2018 ApJ, 852, 38P

Reboussin, L.; Wakelam, V.; Guilloteau, S.; et al. 2015 A&A, 579, 82R

Ricci, L.; Testi, L.; Natta, A.; Brooks, K. J. 2010 A&A, 521, 66R (Ri10)

Schaefer, G. H.; Prato, L.; Simon, M. 2018 AJ, 155, 109S

Simon, M.; Schaefer, G. H.; Prato, L.; Ruiz-Rodriguez, Dary; Karnath, N.; Franz, O. G.; Wasserman, L.H, 2013 ApJ, 773, 28

Simon. M.; Toraskar, J. 2017, ApJ, 841, 95

Simon, M.; Guilloteau, S.; Di Folco, E.; et al. 2017, ApJ, 884, 158S

Somers, Garrett; Pinsonneault, Marc H. 2015, ApJ, 807, 174

Villebrun, F., Alecian, E., Hussain, G. et al., 2019, A&A, 622, 72